\documentclass[preprintnumbers,nofootinbib,amsmath,amssymb,twocolumn]{revtex4}
\pdfoutput=1

\def\ga{\mathrel{\raise.3ex\hbox{$>$\kern-.75em\lower1ex\hbox{$\sim$}}}}
\def\la{\mathrel{\raise.3ex\hbox{$<$\kern-.75em\lower1ex\hbox{$\sim$}}}}

\newcommand\beq{\begin{equation}}
\newcommand\eeq{\end{equation}}
\newcommand\beqar{\begin{eqnarray}}
\newcommand\eeqar{\end{eqnarray}}
\usepackage{graphicx}
\usepackage{epsfig}
\usepackage{epsf}
\usepackage{rotating}
\usepackage{verbatim}

\usepackage{multirow}
\usepackage{natbib}
\usepackage{amsmath}
\usepackage{url}
\usepackage{hyperref}

\newcommand{\bicep}{{\sc Bicep2}}

\newcommand{\wwp}{{\sc WP}}
\newcommand{\WMAP}{{\sc WMAP}}

\newcommand{\planck}{{\sc Planck}}

\preprint{Imperial/TP/2014/CC/2}

\begin{document}

\title{BICEP's acceleration}

\author{Carlo~R.~Contaldi} \email{c.contaldi@imperial.ac.uk}
\affiliation{Theoretical Physics, Blackett Laboratory, Imperial College, London, SW7 2BZ, UK}
\affiliation{Canadian Institute of Theoretical Physics, 60 St. George
  Street, Toronto, M5S 3H8, On, Canada}

\begin{abstract}
  The recent \bicep\ \cite{bicep2} detection of, what is claimed to be
  primordial $B$-modes, opens up the possibility of constraining not
  only the energy scale of inflation but also the detailed
  acceleration history that occurred during inflation. In turn this
  can be used to determine the shape of the inflaton potential
  $V(\phi)$ for the first time - if a single, scalar inflaton is
  assumed to be driving the acceleration. We carry out a Monte Carlo
  exploration of inflationary trajectories given the current
  data. Using this method we obtain a posterior distribution of
  possible acceleration profiles $\epsilon(N)$ as a function of
  $e$-fold $N$ and derived posterior distributions of the primordial
  power spectrum $P(k)$ and potential $V(\phi)$. We find that the
  \bicep\ result, in combination with \planck\ measurements of total
  intensity Cosmic Microwave Background (CMB) anisotropies, induces a
  significant feature in the scalar primordial spectrum at scales
  $k\sim 10^{-3}$ Mpc$^{-1}$. This is in agreement with a previous
  detection of a suppression in the scalar power \cite{Contaldi}.
\end{abstract}

\maketitle

\section{Introduction} The recent \bicep\ detection \cite{bicep2} of
curl patterns, apparently primordial in nature, in the CMB
polarisation pattern, ushers in a new era in the search for a
``complete theory'' of the early Universe. Curl or $B$-type modes can
only be induced in the polarisation by either tensor modes - a
background of gravitational waves present at last scattering, or by
lensing due to structure along the line of sight that distorts
gradient modes ($E$-type) into curl modes. The only other possible
source of $B$-modes, unless new physics is invoked, is foreground
contamination but \bicep\ claims to have ruled out this possibility
with some level of confidence\footnote{However see for example
  \cite{flauger,seljak} for a critical review of the significance of
  the result.}.

In turn, a confirmation that a relic background of super horizon
scaled gravitational waves was present at last scattering will lend
very strong support to the idea that an epoch of inflation occurred in
the very early Universe. A gravitational wave background is a nearly
unavoidable consequence of a period of quasi de Sitter expansion in
the early Universe if it was driven by a single scalar field
\cite{Grishchuk:1974ny,Starobinsky:1979ty,Rubakov:1982df,Fabbri:1983us,Abbott:1984fp}. The
amplitude observed by \bicep\ is close to that predicted by simple
inflationary models such as chaotic
inflation~\cite{Linde:1983gd} and natural
inflation~\cite{Freese:1990rb}.

On the other hand the tensor-to-scalar
ratio $r=0.2$ that fits best the \bicep\ data is in tension with
constraints arising from total intensity CMB measurements
\cite{wmap,planck} on large angular scales. The total intensity on
scales larger than the sound horizon at last scattering is a sum of
both scalar and tensor contributions. A measurement of the total and
fits to the $\Lambda$CDM model with power law primordial spectra
results in a limit on $r$ with $r<0.11$ at 95\% confidence. This
has led to some speculation of non-standard behaviour at large scales
\cite{Contaldi} and/or the requirement of running of
the spectral index and non-standard physics on smaller
scales (see for example \cite{smith}).

In \cite{Contaldi} we noted that a simple suppression of the scalar
power on scales $k\sim 10^{-3}$ Mpc$^{-1}$ easily achieves a
reconciliation of the \bicep\ and \planck\ spectra without creating
additional tension with other small scale probes. This is in contrast
to the addition of curvature to the primordial spectrum, also known as
running of the spectral index. This has two disadvantages; the
first is that the required running is some two orders of magnitude
larger than what is expected in the simplest models of
inflation. The second is that the addition of a curvature term modifies the
spectrum on all scales and will lead to tension with observations on
smaller scales from Large Scale Structure (LSS) surveys.

The required suppression, some 25\% in power, can be achieved in
simple modifications of the single inflaton field paradigm such as the
Starobinsky model \cite{starobinski} as shown in \cite{Contaldi}. Here
we take a model independent approach to analysing the data by
considering the parametrisation of the acceleration history during
inflation as modes observable today were exiting the horizon. This is
in contrast to the standard method that assumes $d\ln P(k)/d\ln k$ is
well described by a Taylor expansion involving the spectral index
$n_s$, a running $d n_s/d\ln k$, etc. or methods parametrising whole
classes of inflation models (see for example \cite{barranco}).

Given a suitable parametrisation of the acceleration we can then
obtain its posterior distribution with respect to observations via
Monte Carlo exploration of the data likelihoods. This in turn allows
us to derive constraints on the primordial spectra and the
inflationary potential if we make model dependent assumptions relating
the power spectra to the original inflaton field perturbations.

This {\sl paper} is organised as follows. In section~\ref{sec:method}
we describe the formalism for generating observables via the
parametrisation of the acceleration occurring during inflation. In
section~\ref{sec:sampling} we describe the sampling method used in our
MCMC exploration of the likelihoods using the parametrised
acceleration formalism. In section~\ref{sec:res} we show the results
of various runs and compare to conventional fits to the data generated
by parametrising the primordial power spectra as power laws and power
laws with running of the spectral index. We obtain posterior distributions for
the acceleration trajectories and its derived corollaries - the scalar
primordial spectrum and inflationary potential. We discuss our results
in section~\ref{sec:disc}.

\section{Method}\label{sec:method} 

Assuming a Friedmann-Robertson-Walker background
with scale factor $a(t)$ evolving with cosmological time $t$ the
background evolution is described completely by the Hubble rate
$H=\dot a/a$ and the quantity $\epsilon$ which is a measure of the
acceleration in $a$
\begin{equation}
  \epsilon = -\frac{\dot H}{H^2}\equiv 1-\frac{\ddot a}{a}\,,
\end{equation}
where $H=\dot a/a$ and overdots represent derivatives with respect to
time $t$. By definition when $\epsilon < 1$ the rate of change of $a$
is increasing and the Universe is inflating. When $\epsilon \ll 1$ the
Hubble rate is nearly constant and the background is close to de
Sitter or in the ``slow-roll'' regime as long as higher derivatives of
$H$ are also small.

For convenience it is useful to switch independent variable to $e$-folds
defined as $N=\ln a/a_0$. Given a background history or
``trajectory'' in $H(N)$ and  $\epsilon(N)$ we can define observable
quantities. For example, to first order in the slow-roll expansion
\cite{lyth}, the resulting primordial curvature and
tensor
mode dimensionless spectra are given by
\begin{equation}\label{eq:scalar}
  k^3P_s(k) =\left. A_s\,\frac{H^2(N)}{\epsilon(N)}\right|_{N=N_k}\,,
\end{equation}
and
\begin{equation}\label{eq:tensor}
  k^3P_t(k) =\left. 16\,A_s\,H^2(N)\right|_{N=N_k}\,,
\end{equation}
respectively. Here $N_k$ is the $e$-fold at which mode with fourier
wavenumber $k$ exits the
horizon with $k=aH$ and $A_s$ is the primordial normalisation of the
perturbations. The ratio of the two spectra 
\begin{equation}
  r(N) \equiv  \frac{P_t}{P_s} = 16\, \epsilon(N)\,,
\end{equation}
is often quoted, at a chosen pivot scale, as the model defining
observable as in the \bicep\ case. 

\begin{table*}
  \centering                         
  \begin{tabular}{|l|c|l|}       
    \hline\hline                
    Parameter & Prior range & Definition \\    
    \hline                        
    $\omega_b\equiv \Omega_b\,h^2$ & [0.005,0.1]& Baryon density today\\
    $\omega_c\equiv \Omega_c\,h^2$ & [0.001,0.99]& Cold dark matter density today\\
    $100\, \theta_{MC}$ & [0.5,10.0]& 100 $\times$ {\tt CosmoMC} sound horizon to angular diameter distance ratio approximation\\
    $\tau$ & [0.01,0.8]& Optical depth to reionisation\\
    $\ln(10^{10}A_s)$ & [2.7 4]& Scalar spectrum normalisation\\
    $\ln\epsilon_i$ &[-11.5,-1.6]& Number of $e$-folds for which trajectory is integrated back from end of inflation\\[0.1cm]
    \hline
    $n_s(k_\star)$&... & Scalar spectral index measured from trajectory spectrum at scale $k_\star=0.05$ Mpc$^{-1}$\\
    $r(k_\star)$& ...& Tensor-to-scalar ratio measured from trajectory spectra at scale $k_\star=0.05$ Mpc$^{-1}$\\
[0.1cm]
    \hline                                   
  \end{tabular}
 \caption{Uniform MCMC priors for cosmological parameters and their
    descriptions. \planck\ nuisance parameters are not listed here but
    are included with the same prior settings as used in
    \cite{planck_pars}. The second block are derived
    parameters that are not sampled directly.}  
  \label{tab:params}  
\end{table*}

Both these spectra source anisotropies in the total intensity of the
CMB but only the tensor spectra sources the $B$-modes of the
polarisation. In both cases the tensor contribution is negligible
beyond multipoles $\ell\sim 200$ as gravitational waves decay once
they re-enter the horizon and only modes that were larger than the
horizon scale at recombination affect the pattern of
anisotropies.

Assuming a single scalar field $\phi$ rolling down a potential
$V(\phi)$ is driving inflation the Hubble equation is given by
\begin{equation}\label{hub}
H^2 = \frac{1}{3M_{\rm pl}^2} \left(\frac{1}{2}\dot\phi^2+V(\phi) \right)\,,
\end{equation}
where $M_{\rm pl}$ is the reduced Planck mass. The inflaton
equation of motion is
\begin{equation}
\ddot \phi +3H\dot\phi+\frac{\partial V}{\partial \phi}=0\,.
\end{equation}
These can be combined to relate $\epsilon$ to the time derivative of
the inflaton
\begin{equation}\label{eq:epsiphi}
  \epsilon = \frac{1}{2M_{\rm pl}^2}\left( \frac{\dot\phi}{H}\right)^2\,.
\end{equation}
The above can then be substituted back into (\ref{hub}) to obtain a
relationship between $\epsilon$, $H$ and the scalar potential
\begin{equation}\label{potential}
  V[\epsilon(N)] = 3M^{2}_{pl}H^{2}(N)\left[ 1 - \frac{\epsilon(N)}{3}\right]\,.
\end{equation}
In turn this can be used to reconstruct $V[\phi(N)]$ by relating
$\phi$ to $\epsilon(N)$ via (\ref{eq:epsiphi})
\begin{equation}\label{eq:phi}
  \phi(N) - \phi_0 = -\sqrt{2}\int^N_0 \sqrt{\epsilon(\tilde
    N)}\,d\tilde N\,.
\end{equation}

At any time ($e$-fold) the Hubble rate can be obtained, up to a normalisation, by integrating
$\epsilon$ too 
\begin{equation}
  \ln H(N) -\ln H_0 = - \int^N_0 \epsilon(\tilde
  N)\,d\tilde N\,.
\end{equation}
Similarly the scale of the mode exiting the horizon at each $e$-fold
can be calculated using $k=aH$
\begin{equation}\label{eq:k}
  \ln k(N) -\ln k_0= N + \ln H(N) = N - \int^N_0 \epsilon(\tilde
  N)\,d\tilde N\,.
\end{equation}
The implicit assumption made in~(\ref{eq:k}) is that we know the exact
number of $e$-folds that occurred after the Universe stopped
inflating. This is sensitive to the exact evolution of the background
during the reheating epoch which would shift the relation between $\ln
k$ and $N$ systematically by a few $e$-folds. We will ignore this
unknown shift when relating $e$-folds to observable scales.

Thus, assuming first order in slow-roll, all quantities required to
compute observables can be calculated by specifying the function
$\epsilon(N)$ and a normalisation scale $A_s$. 

\section{Sampling the acceleration}\label{sec:sampling}

In our method the functional space $\epsilon(N)$ is the basis for the
Markov Chain Monte Carlo (MCMC) exploration of data likelihoods. We
sample the space by drawing random amplitudes in $\ln \epsilon$ at a
series of regularly spaced spline $e$-folds $N_i$. The spline points
are then used to reconstruct the full function $\epsilon(N)$ using a
cubic spline \cite{numrec}.

A cubic spline ensures a sufficiently smooth function for our purposes
and the sampling of $\ln \epsilon$ ensures that we always obtain an
inflating solution with $H$ decreasing monotonically with time or
$e$-fold. Our use of the first order slow roll approximation also
means that we only need to integrate the splined function to obtain
the spectra and scale definitions. If higher orders or the full, numerical
solutions, were required the derivatives of the function $\epsilon(N)$
would have to be defined. This would require a higher order
interpolation scheme and more parameters to have $d^2\epsilon/dN^2$ be
continuous and compute quantities to second order in slow roll.

First order in slow roll may seem restrictive in the context of
exploring structure in the inflationary trajectories but the data is
suggesting a transition between two regimes that are well within the
slow roll approximation \cite{Contaldi} and this should therefore be
sufficient to define the trajectories in the observational window. 

We use the {\tt CosmoMC} package \cite{cosmomc} for the MCMC
exploration and modify it to search in the space of spline point
amplitudes $\epsilon_i$ where $i=1,...,N_{\rm sp}$ instead of the usual
primordial parameters $n_s$, $r$, $dn_s/d\ln k$, $n_t$, etc. The
spline points are positioned regularly between $N=0$ and $N=N_{\rm
  max}$ corresponding to the largest and smallest modes $k$ required by
the {\tt CAMB} package \cite{camb} in order to compute the CMB spectra
between multipoles $2\leq \ell \leq 2500$. The range of scales required,
typically ${\cal O}(10^{-5})\leq k \leq {\cal O}(10^{-1})$, extends
well beyond the range where CMB observations have a significant impact
but are required in order to integrate the radiation transfer
functions to a sufficient degree of accuracy. Thus the regular spacing
in such a wide range is not optimal but it avoids the need to
extrapolate the functions beyond the range covered by our splining. 

At each MCMC sample we compute the cubic spline of the $\epsilon_i$
amplitudes to obtain a continuous function $\epsilon(N)$ in the chosen
range. This function is integrated to obtain $H(N)$, the corresponding
scales $\ln k(N)$, and the inferred value of the field $\phi$. The
scalar and tensor primordial power spectrum are then calculated
using~({\ref{eq:scalar}) and~(\ref{eq:tensor}) respectively. These are
  then passed onto {\tt CAMB} for convolution with the CMB radiation
  transfer functions.

The cubic spline method requires an assumption for the derivatives at
the boundaries and we adopt the ``natural spline'' assumption by
imposing that the curvature vanishes at the boundaries i.e. that the
gradient is constant.

\begin{figure*}
  \centering
  \includegraphics[width=6in]{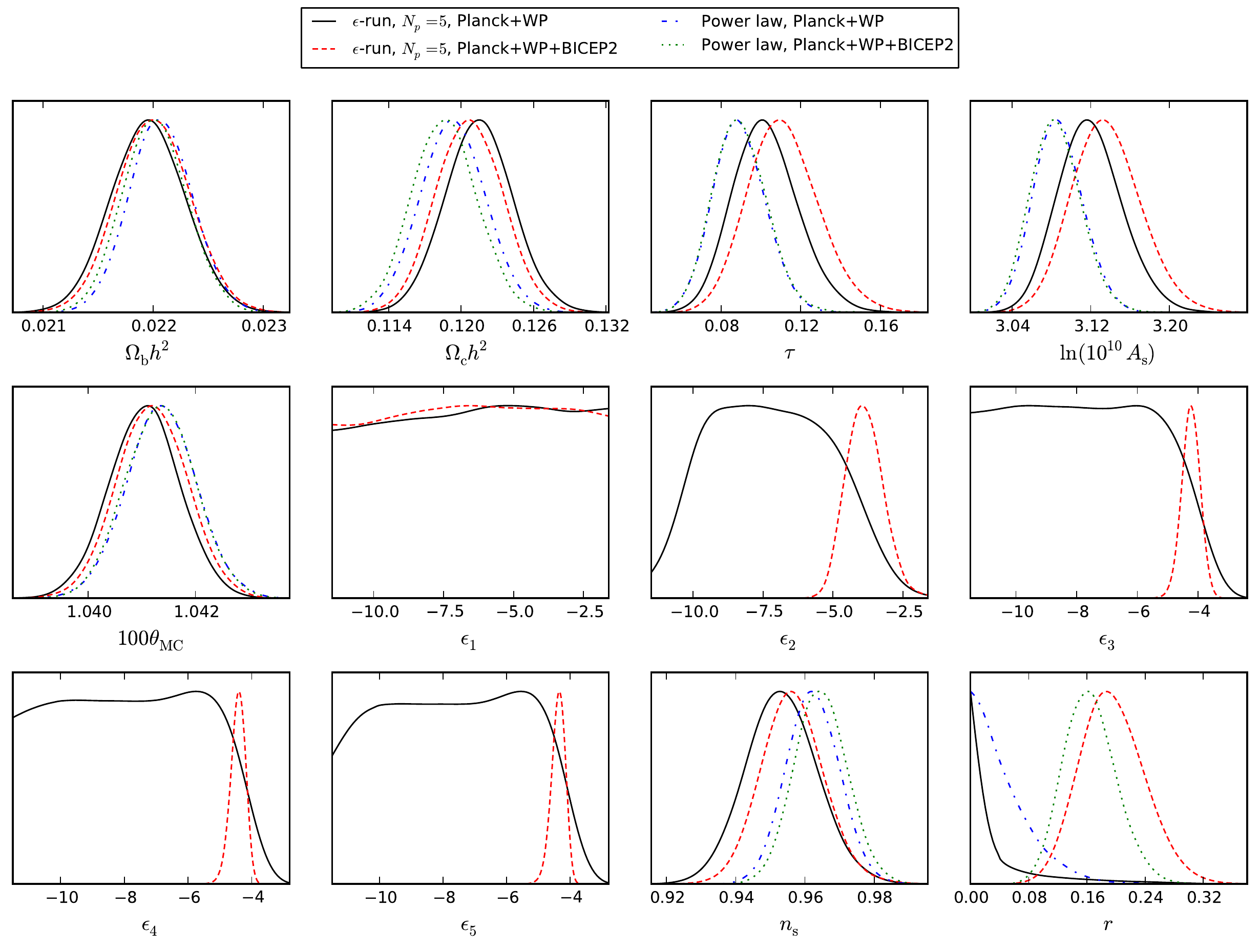}
  \caption{1-d marginalised posteriors for the ten basic and two ($n_s$
    and $r$) derived parameters. The $\epsilon$ runs are all for
    $N_{\rm sp}=5$ and are compared to the power law model runs for
 data combinations including and excluding \bicep. In the no \bicep\
 case the $\epsilon$ spline point amplitudes are not constrained
 although the significant correlation between each spline point results
in functions with tilts that are still compatible with observations.}
  \label{fig:epsilon5}%
\end{figure*}

We run {\tt CosmoMC} as setup for \planck\ data runs. These include a
number of \planck\ nuisance parameters that are used to marginalise over
systematic or astrophysical residuals in the \planck\ spectra. The data
combination also includes \WMAP\ polarisation and we label this
``\planck+\wwp''. In addition to this we include the \bicep\ results as
provided in the most recent {\tt CosmoMC} release. Uniform priors for
all nuisance parameters are left unchanged from the standard \planck\ runs. 

Table~\ref{tab:params} gives a summary of the uniform priors assumed
for cosmological parameters and lists the definitions of the derived
parameters $n_s$ and $r$ used to compare with conventional runs
assuming power law spectra. The derived parameters are calculated
using the first order slow-roll approximations
\begin{eqnarray}
  n_s &=& 1-4\epsilon+2\eta\,,\nonumber\\
  r &=& 16\epsilon\,,\\
  n_t &=& -2\epsilon\,,\nonumber
\end{eqnarray}
where $\eta=\epsilon-d\ln\epsilon/dN$ and is calculated analytically
from the cubic spline for $\ln\epsilon$.

The MCMC chains are run until the $R^{-1}$ convergence parameter is
smaller than 0.01. We run cases with $N_{\rm sp}$=5, 6, and 7. For
$N_{\rm sp}>7$ the MCMC sampling becomes very inefficient and the
required convergence criterion is difficult to achieve due to the
large correlations between the $\epsilon_i$ spline point amplitudes. A
typical run with eight parallel chains contains several hundred
thousand accepted models. The minimal case, with $N_{\rm sp}=5$, includes two
boundary points and three internal points so should be able to fit the
data at least as well as spectral models with three or four parameters
e.g. $n_s$, $r$, $dn_s/d\ln k$, and an (independent) $n_t$.

\begin{figure*}
  \centering
  \includegraphics[width=6in]{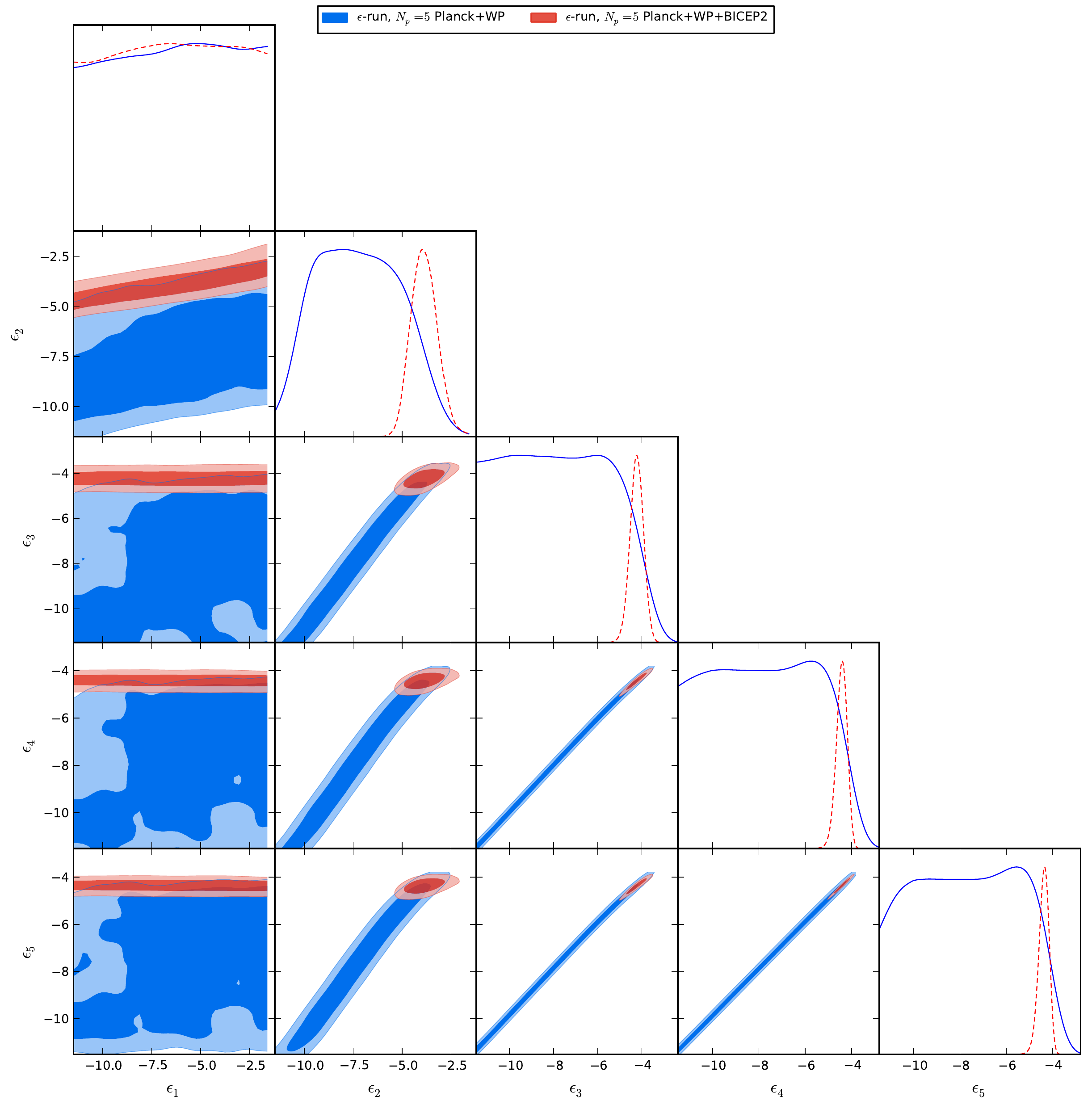}
  \caption{Triangle plot showing the 2-d marginalised posteriors for
    the spline point amplitudes in the $N_{\rm sp}$ run. The contours
      indicate 68\% and 95\% confidence regions and we compare the
      case with and without \bicep\ included. When \bicep\ is included
    the spline point amplitudes are tightly constrained although the
    smaller scale points $\epsilon_4$ and $\epsilon_5$ are very
    correlated. This is driven by high-$\ell$ CMB observations tightly
    constraining the allowed gradient on small scales.}
  \label{fig:epsilon5_tri}%
\end{figure*}

\section{Results}\label{sec:res}

\subsection{Acceleration}

\begin{table}[t]
\centering
\begin{tabular}{|l| c |c c|c c|}
 \hline 
 \hline
  Model &Data & $\Delta N_p$  & $\Delta \chi^2$&$\Delta\mbox{\sl AIC}$&Rel. Like\\[0.1cm]
  \hline
  Power law& {\sc P}+{\sc WP} & 0 & -- &-0.066&0.968\\
  $\epsilon$ $N_{\rm sp}=5$& {\sc P}+{\sc WP}& +3  & -6.070&--&1.000 \\
  \hline
  Power law& {\sc P}+{\sc WP}+{\sc B2} & 0  & --&-4.920&0.085 \\
  Running& {\sc P}+{\sc WP}+{\sc B2} & +1  & -6.68&-2.42&0.785 \\
  $\epsilon$, $N_{\rm sp}=5$& {\sc P}+{\sc WP}+{\sc B2} & +3  & -10.92&--&1.000\\
  $\epsilon$, $N_{\rm sp}=6$& {\sc P}+{\sc WP}+{\sc B2} & +4  & -12.50&-0.422&0.810
  \\
  $\epsilon$, $N_{\rm sp}=7$& {\sc P}+{\sc WP}+{\sc B2} & +5  & -12.22&-2.704&0.259 \\
  \hline
\end{tabular}
\caption{Best-fit $\chi^2$ values for the various $\epsilon$ runs
  compared to standard, power law runs with ({\sc P}+{\sc WP}+{\sc B2}
 ) and without \bicep\ ({\sc P}+{\sc WP})
  data. $\Delta N_p$ is the change in total number of model parameters
  compared to the power law case which has $N_p=21$ (7 cosmological + 14
  nuisance) parameters and the case including running of the spectral index. The relative likelihood is evaluated using the
  Akaike Information Criterion (AIC) with respect to the model in both data
  combination that gives the lowest AIC value. We see that when \bicep\
  is absent the $\epsilon$ model is not favoured significantly over
  the power law model. When \bicep\ is included the power law model has
  a relatively likelihood of only 8.5\% with respect to the $N_{\rm
    sp}=5$ $\epsilon$ model. We can also conclude that the $N_{\rm
    sp}=5$ model is slightly more motivated than the power law + running
  case i.e. it is just as motivated by the data as a solution to the tension.}
\label{tab:chi2}
\end{table}

\begin{figure}[t]
  \centering
  \begin{tabular}{c}
  \includegraphics[width=3.5in]{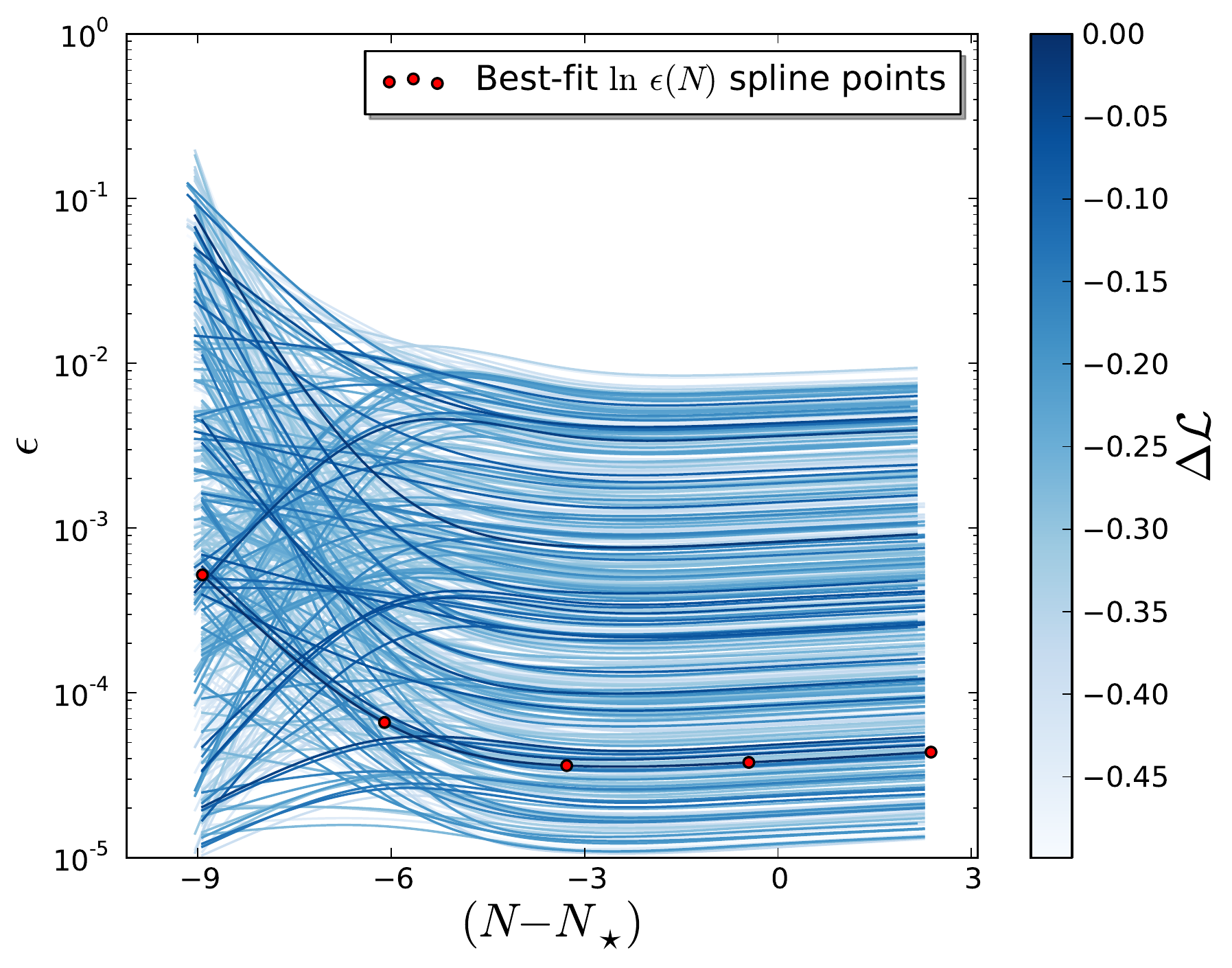}\\
  \includegraphics[width=3.5in]{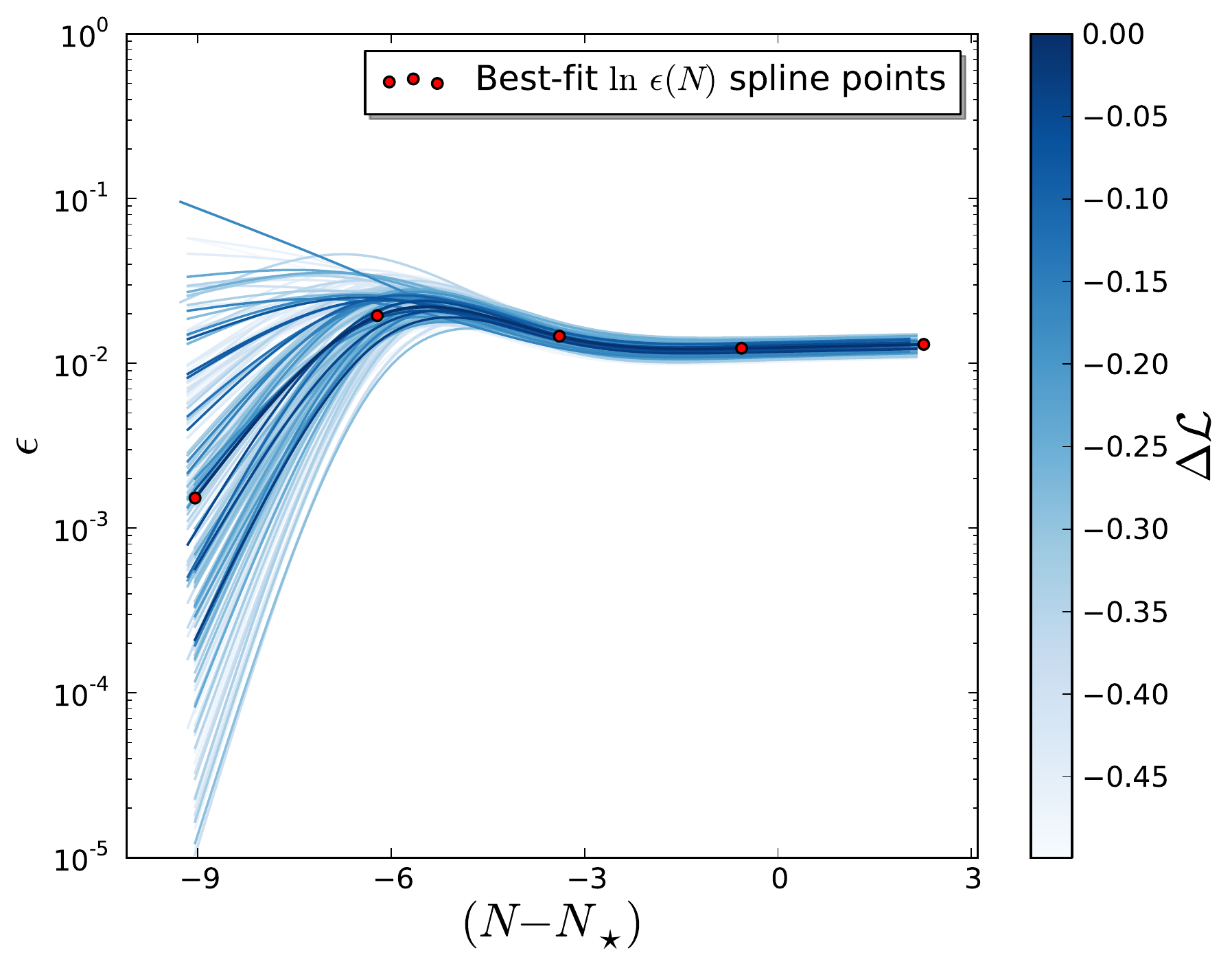}
  \end{tabular}
  \caption{Ensemble of splined trajectories that lie within $|\Delta {\cal
      L}|<0.5$ from the best-fit model in the $N_{\rm sp}=5$ case
    excluding ({\sl top}) and including ({\sl bottom}) the \bicep\
    data. Each
  $\epsilon$ trajectory is colour coded by distance from the
  best-fit. The (red) points show the spline point amplitudes for the
  best-fitting trajectory. The range in $e$-folds is shifted with
  $N_\star$ corresponding to the $e$-fold at which the scale $k=0.05$
  Mpc$^{-1}$ exited the horizon.}
  \label{fig:epsilon5_lines}%
\end{figure}

We start by looking at the minimal $N_{\rm sp}=5$ run. In
Figure~\ref{fig:epsilon5} we show the marginalised posterior
distribution for the 10 cosmological parameters and two derived
parameters, $n_s$ and $r$. We compare the \planck+\wwp\ and
\planck+\wwp+\bicep\ data combinations for both $\epsilon$ and power
law runs. The $\epsilon$ spline point amplitudes are largely
unconstrained when \bicep\ is not included. Despite this the posterior
in $n_s$ is well constrained in the $\epsilon$ runs. This is due to
the fact that although the amplitude is unconstrained for any
particular trajectory it is highly correlated with its gradient
($\eta$) and the combination is fixed by the data. When \bicep\ is
included all spline points except the first one are well
constrained. This is due to the measurement of tensor mode amplitude
which fixes $\epsilon$ directly.

The posterior in $r$ is consistent with zero in both cases without
\bicep\ data. The difference in shape of the posterior is driven by
our $\ln \epsilon \equiv \ln r$ prior {\sl versus} the linear prior
used in the conventional power law run. When \bicep\ is included the
posteriors are similar and the choice of prior is not a dominant factor
anymore since the detection is significant. However is does affect the
location of the peak in the posterior and that of correlated variables
$\tau$ and $A_s$. 

The correlations between spline point amplitudes in the two $\epsilon$
runs are shown in Figure~\ref{fig:epsilon5_tri}. We see that the
spline points at higher $N$, corresponding to modes exiting the horizon
later in inflation, are highly correlated due to the tightly
constrained tilt of the CMB anisotropies at high-$\ell$ as observed by
\planck. This also explains why the last spline point amplitude,
corresponding to a scale of $\sim 0.1$ Mpc$^{-1}$ is apparently well
constrained despite the fact that CMB observations by \planck\ do not
extend to those scales.

Figure~\ref{fig:epsilon5_lines} shows the ensemble of splined
$\epsilon$ trajectories in the $N_{\rm sp}=5$ for the \planck+\wwp\
and \planck+\wwp+\bicep\ data combinations. The trajectories are
selected from the MCMC chains to be within 0.5 in log likelihood
${\cal L}$ from the best-fitting trajectory. This would be equivalent
to the 1-$\sigma$ set if the likelihood were Gaussian. There are 250
trajectories within this range of $\Delta{\cal L}$ for the accepted
MCMC steps in the chain. Each curve is colour coded according to the
$\Delta{\cal L}$ value of its fit to the data
(\planck+\wwp+\bicep). We see that the shape of $\epsilon(N)$ is well
constrained in both cases (with and without \bicep). However its
amplitude is not well constrained when \bicep\ data is absent since a
direct measurement if $\epsilon$ is not possible unless $B$-modes are
detected. At later times (larger $N$) the combination of \bicep\ and
\planck's high-$\ell$ CMB observations tightly focus
$\epsilon(N)$. The small, positive gradient in this regime
gives rise to the slightly red tilted ($n_s < 1$) scalar spectrum. At
early times the function is essentially unconstrained but this region
does not affect the observables. The rise in amplitude at
$N-N_\star\sim -5$ will cause a suppression of the scalar power since
$H$ is approximately constant and the power is proportional to the
inverse of $\epsilon$. This is the feature which has been noted in the
literature and is driven by the \planck\ {\sl vs} \bicep\ tension.

In Table~\ref{tab:chi2} we show a summary of the model comparisons
between the power law assumption and $N_{\rm sp}=5$, 6, and 7
runs. We quote both $\Delta \chi^2$ values for the best-fitting models
in each MCMC chain and the Akaike Information Criterion (AIC)
assessment of their relative likelihoods \cite{AIC}. 

The {\sl AIC} is defined as  {\sl AIC} $= 2\,N_p + \chi^2$,
where $N_p$ is the number of parameters in the model. It attempts to
properly take into account the penalty for using models with
increasing number of parameters to describe a set of data. The
relative likelihood between a model with minimum {\sl AIC} value and a
second model 
\begin{equation}
  L_{\mbox{\sl AIC}} = \exp\left[(\mbox{\sl AIC}_{\mbox{min}}-\mbox{\sl AIC})/2\right]\,,
\end{equation}
gives an estimate of the probability that the second model minimizes
the information required to describe the data over the first
one. $N_p=21$ for the tensor, power law model run including the \planck\ data
- 7 cosmological parameters and 14 nuisance parameters. 

We see that the $\epsilon$ model with $N_{\rm sp}=5$ uses the minimum
information to describe the data compared to all other models
considered. It is comparable to the power law + running case in this
aspect. However if higher-$\ell$ data or LSS constraints were to be
added the relative likelihood of the running model would drop
significantly due to tight limits on the running from small
scales. Thus the $\epsilon$ model is favoured by the data and this is
driven almost entirely by the \bicep\ data since the likelihoods are
not significantly different in the no \bicep\ case.

The $N_{\rm sp}=6$ performs nearly as well as the $N_{\rm sp}=5$ case
however the $N_{\rm sp}=7$ is considerably disfavoured, it contains
too many parameters to describe the required trajectory and an
oversampled spline can induce too much structure which is not favoured
by the data. We have also attempted a $N_{\rm sp}=4$ run however this
did not converge due to the under-sampled function being dominated by
the uncertainty in the boundary point at low $N$. It may turn out
however, that a non-regularly spaced set of 4 spline points may have
enough degrees of freedom to fit the structure required by
the data. In that case it may well prove to be even more favoured than
the  $N_{\rm sp}=5$ model since it has one less parameter. We leave
this for future work.

\subsection{Power spectra}

\begin{figure}[t]
  \centering
  \includegraphics[width=3.5in]{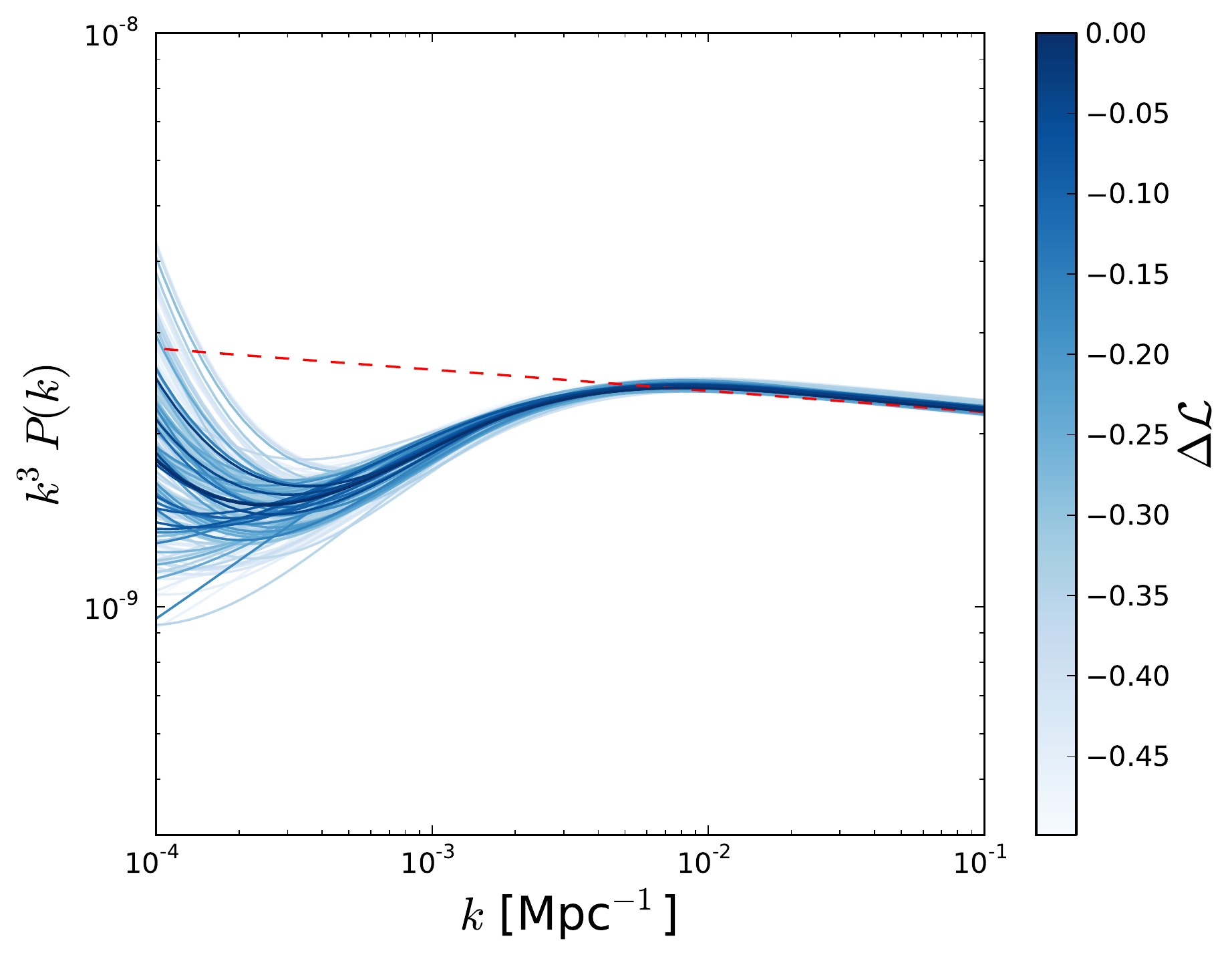}	
  \caption{Similar to Figure~\ref{fig:epsilon5_lines} but for the
    dimensionless scalar power spectrum. The range in $k$ is cutoff on
  the largest scales with respect to Figure~\ref{fig:epsilon5_lines}
  since the first order slow-roll approximation becomes inaccurate for
the the trajectories with the largest $\epsilon$ (smallest
$k^3P(k)$). The suppression of power driven by the \planck\ {\sl vs}
\bicep\ tension is clearly seen when compared to a reference power law
model (dashed/red line) with $n_s=0.962$.}
  \label{fig:power5}%
\end{figure}

\begin{figure*}
  \centering
  \begin{tabular}{ll}
    \includegraphics[width=3.0in,trim=0cm 0cm 0cm
  0cm,clip]{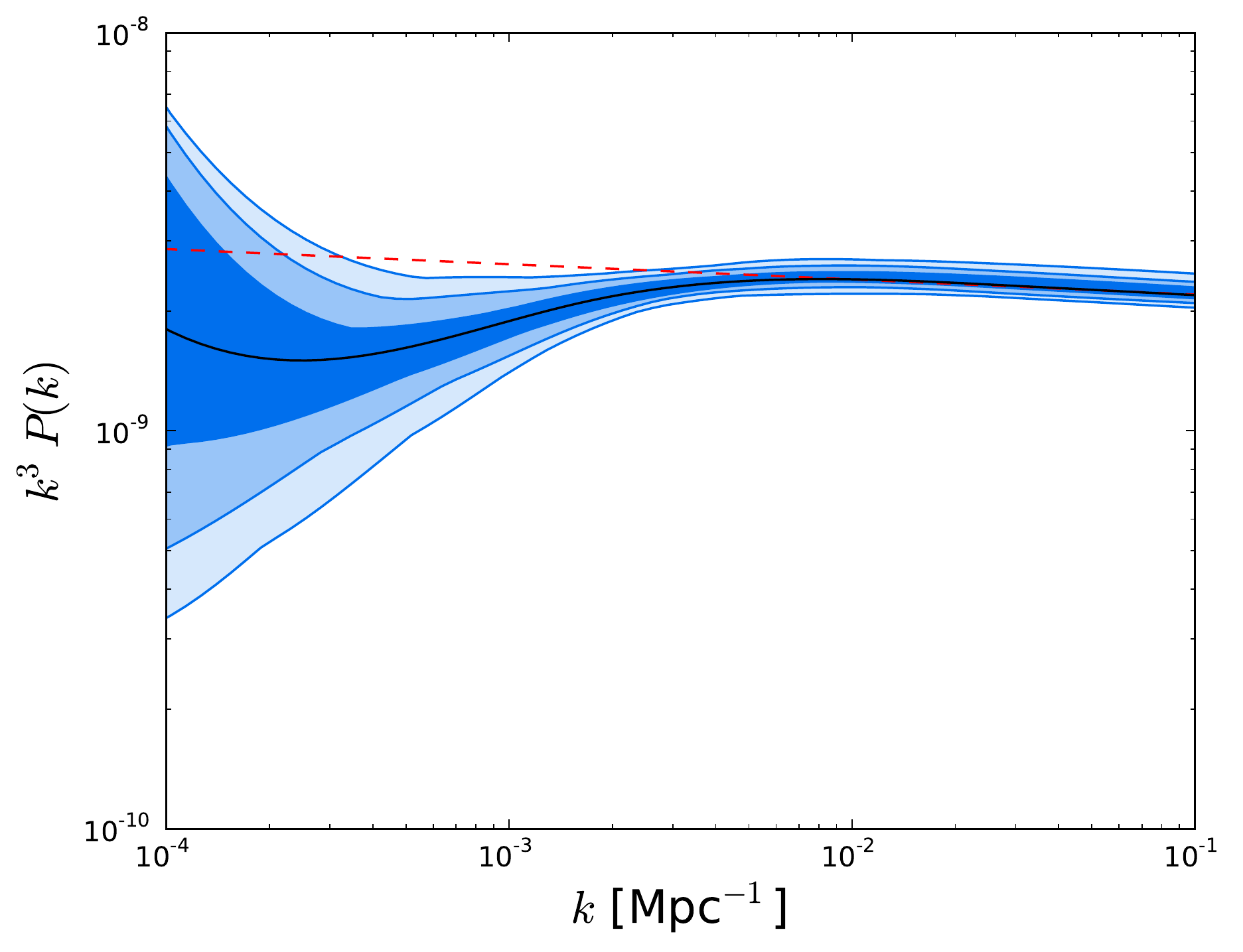}&
    \includegraphics[width=3.0in,trim=0cm 0cm 0cm
  0cm,clip]{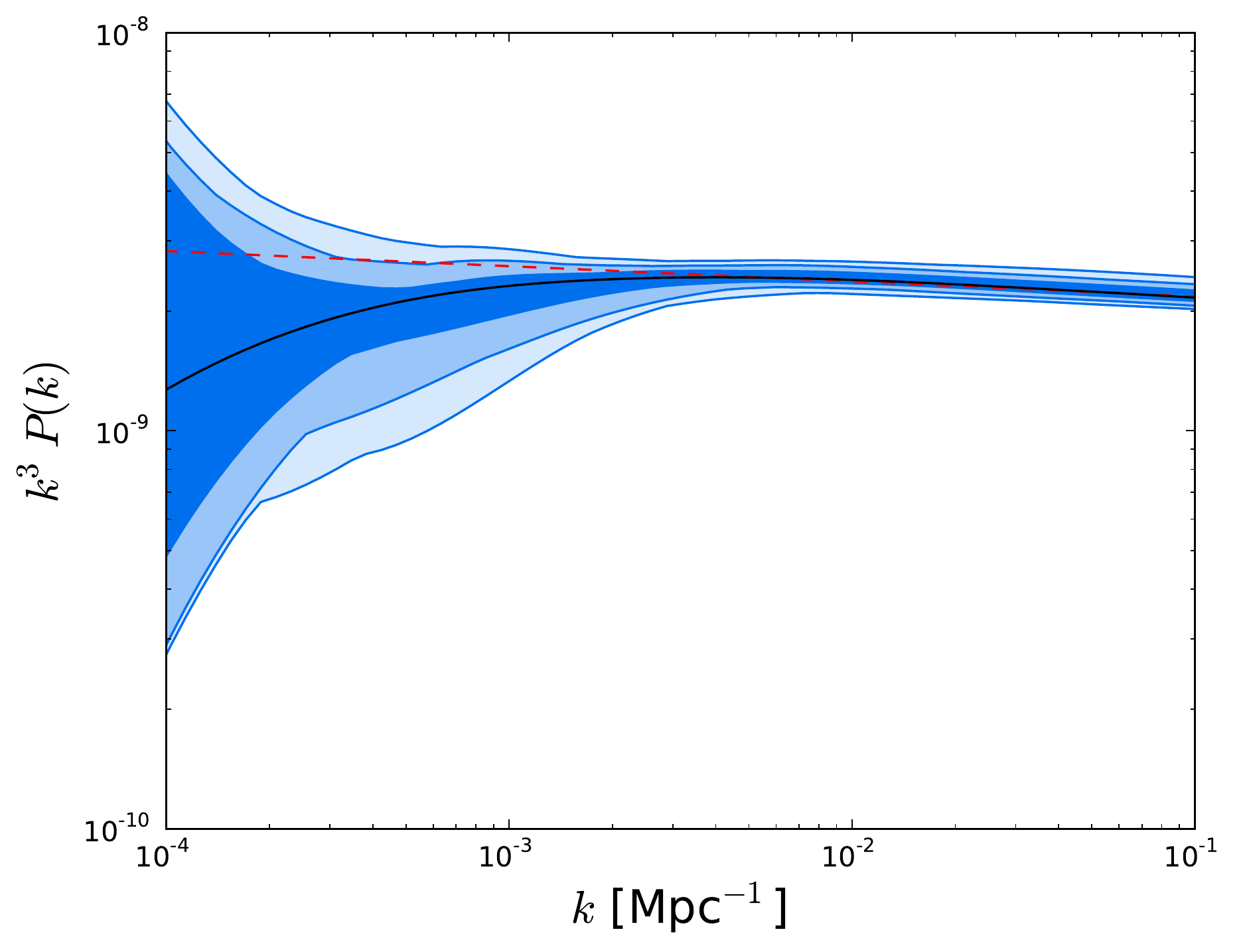}\\
    \includegraphics[width=3.0in]{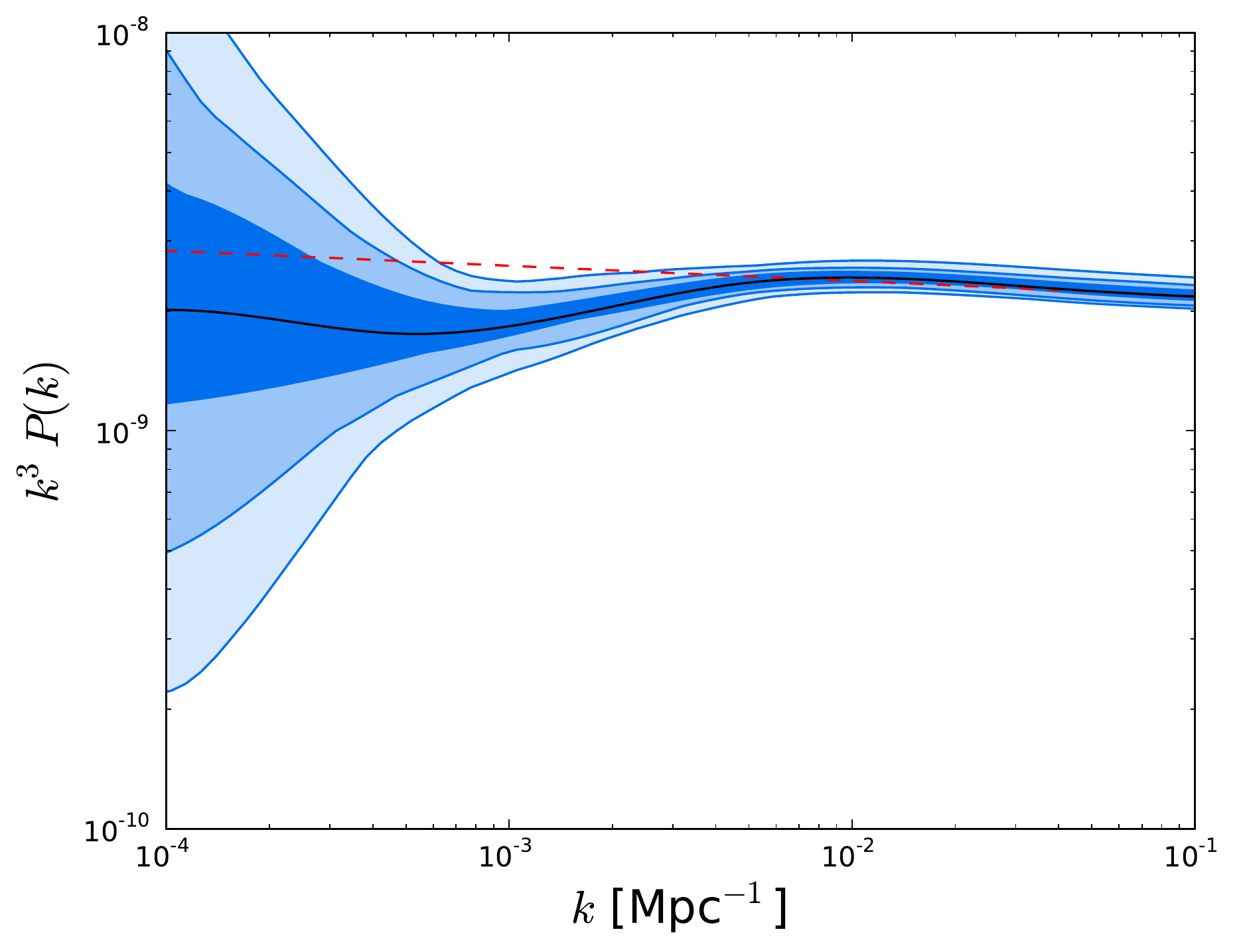}&
    \includegraphics[width=3.0in,trim=0cm 0cm 0cm
  0cm,clip]{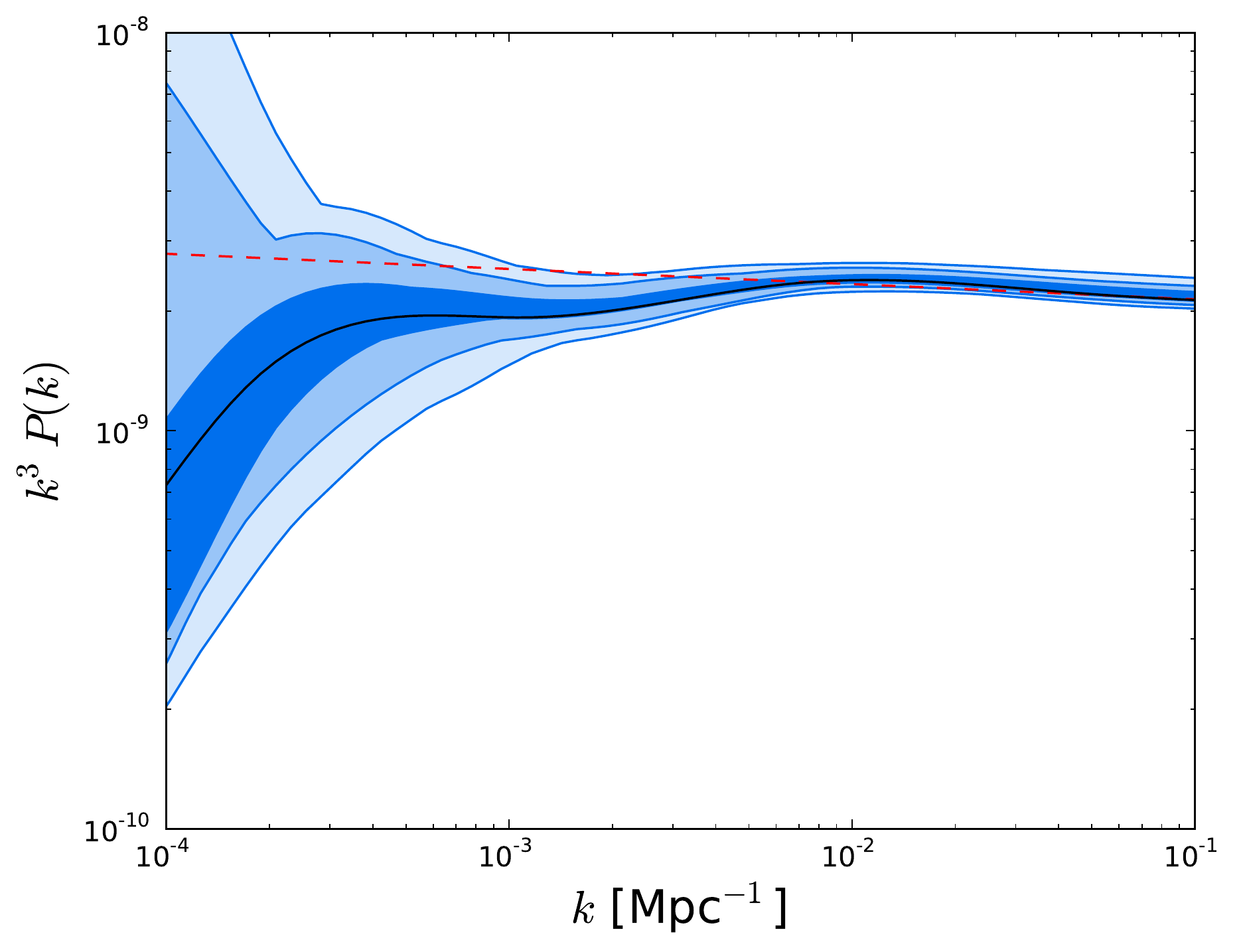}
  \end{tabular}	
  \caption{Confidence levels showing $\Delta {\cal L}=-0.5$, -2.0, and
    -4.5 for ({\sl from top-left, clockwise}): The $N_{\rm sp}=5$ run,
    the  $N_{\rm sp}=5$ with no \bicep\ included, the $N_{\rm sp}=6$,
    and $N_{\rm sp}=7$ runs. In each case the red (dashed) line is a
    reference power law with $n_s=0.962$. The 3-$\sigma$ contours
    encompass the power law over all scales only in the no \bicep\
    case. The $N_{\rm sp}=7$ case is only mildly incompatible with the
    power law case but this model is not favoured by the data.}
  \label{fig:powers}%
\end{figure*}

The significance of the feature, as imprinted on the scalar power
spectrum, can be seen in Figure~\ref{fig:power5} which shows the
scalar spectrum, as a function of $k$, for the same set of
trajectories. The ensemble is compared to a reference power law model
that is compatible with the small scale regime around the pivot scale
$k_\star=0.05$ Mpc$^{-1}$. The reference power law has a spectral
index $n_s=0.959$ compatible with the peak in the posterior for $n_s$
in the run.

The validity of the first order slow-roll assumption used
in~(\ref{eq:scalar}) is limited when $\epsilon\sim 0.1$ for some of the
trajectories close to the lower $e$-fold bound. We therefore restrict
the $k$ range to smaller scales when looking at the resulting
spectra. In the region of the feature of interest $k\sim 10^{-3}$
Mpc$^{-1}$ the value of epsilon converges to ${\cal O}(10^{-2})$ which
leads to a few percent accuracy in the spectra and other observables.

To assess the stability of the suppression feature seen in the scalar
power spectrum we compare similar ensembles for the three different
$\epsilon$ runs in Figure~\ref{fig:powers}. The shadings indicate the
maximum and minimum bounds covered by ensembles of trajectories that
have $\Delta {\cal L}=-0.5$, -2.0, and -4.5 i.e. equivalent to 1, 2,
and 3-$\sigma$ thresholds for a Gaussian likelihood. The no \bicep\
case is also shown for the $N_{\rm sp}=5$ case showing that the
reference power law model is compatible with the indicative 2-$\sigma$
region. For both favoured models with $N_{\rm sp}=5$ and 6 the power
law model is outside of the 3-$\sigma$ region around the scales where
the power is suppressed. In the less favoured $N_{\rm sp}=7$ case the
power law lies just outside the 3-$\sigma$ region for a very limited
range in scales. The analysis indicates that the feature is
significant for the most favoured models, or alternatively that the
power law hypothesis can be ruled out.

In both the $N_{\rm sp}=5$ and $N_{\rm sp}=6$ models favoured by the data the
best-fitting power spectra and distribution around it indicate that
the power either levels off or grows in amplitude going to larger
scales, $k<5\times 10^{-4}$. This is in agreement with the analysis of
\cite{Contaldi} where a mild, step-like suppression was found to be
preferred by the data significantly as opposed to a sharp cutoff due
to a fast-roll to slow-roll transition \cite{cutoff}. In
terms of the acceleration, this indicates that the period of inflation
extended to earlier times with $\epsilon$ remaining less than unity
rather than the feature being due to the start of inflation being just
outside the observational window.

\begin{figure*}[t]
  \centering
  \begin{tabular}{ll}
  \includegraphics[width=3in]{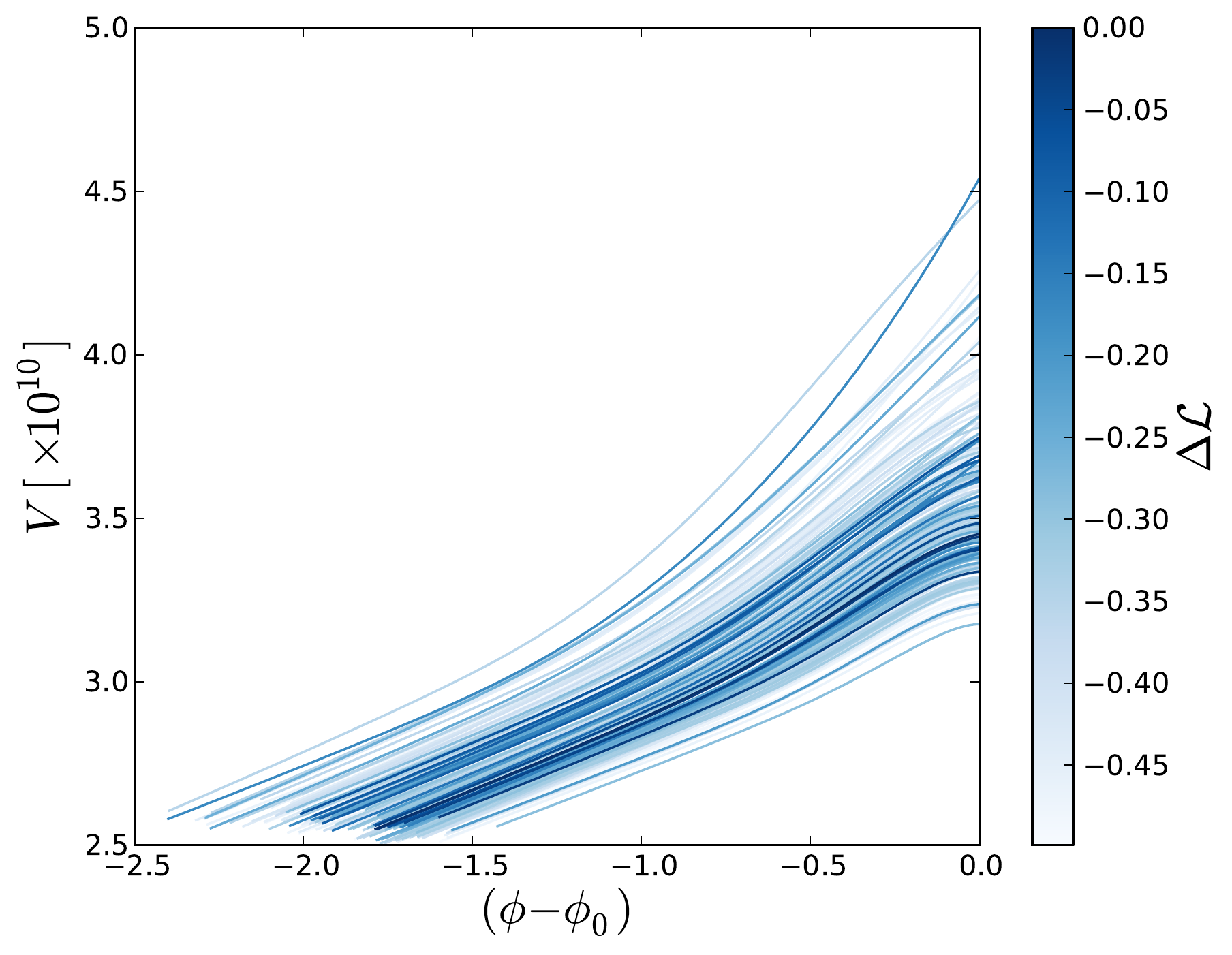}&
  \includegraphics[width=3in]{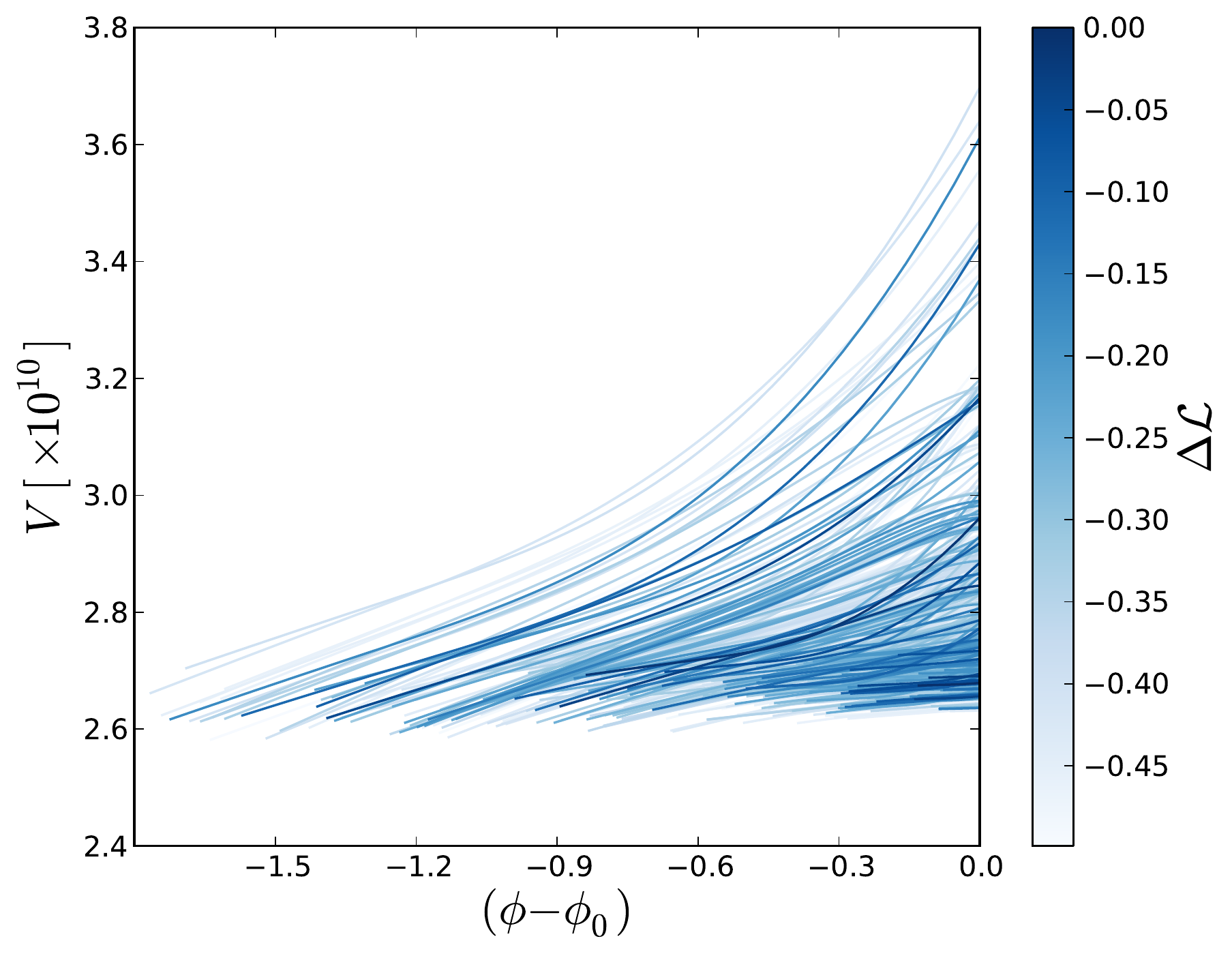}\\
  \includegraphics[width=3in]{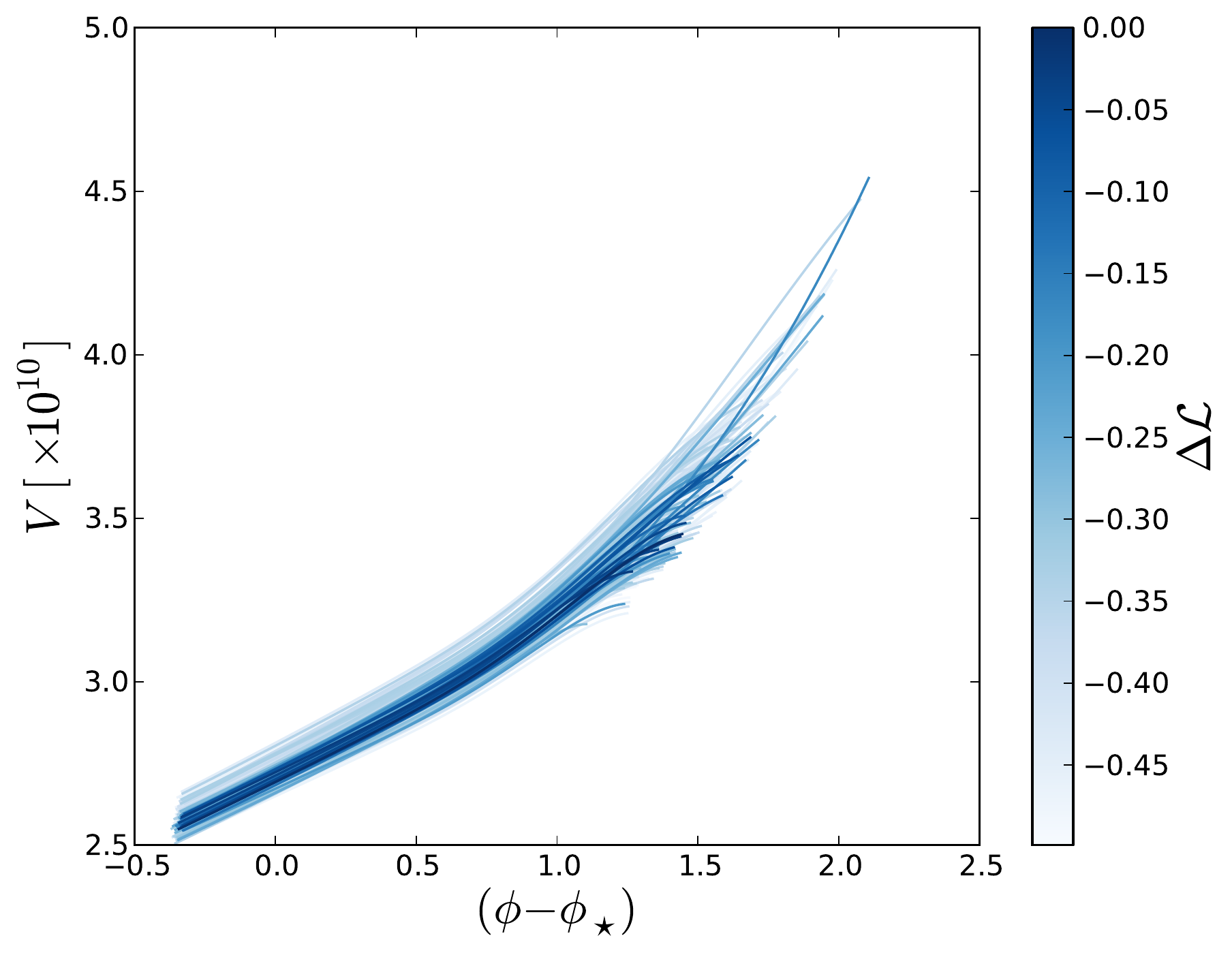}&
  \includegraphics[width=3in]{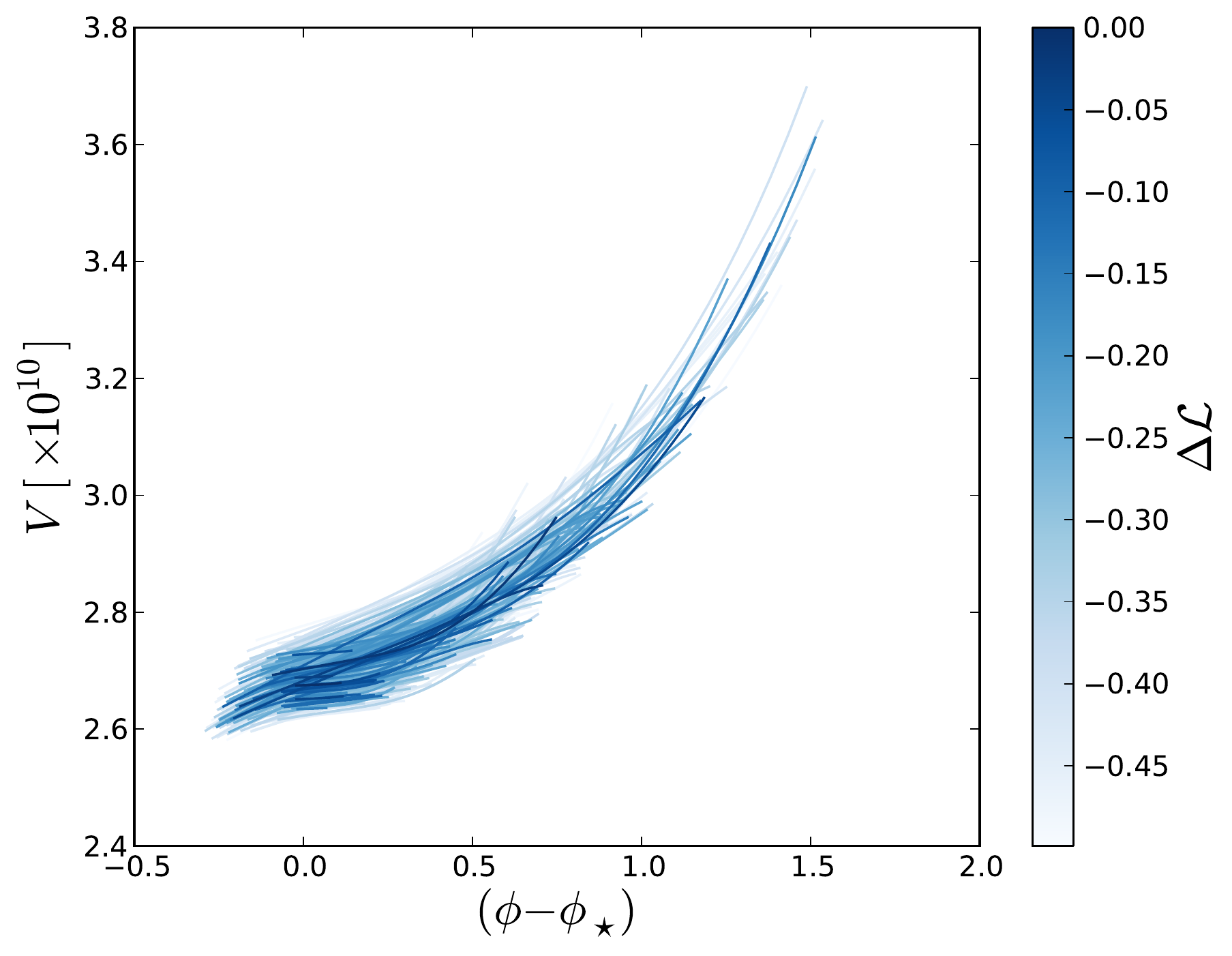}
  \end{tabular}
  \caption{The inflaton potentials $V(\phi)$ corresponding to the
    ``1-$\sigma$'' $N_{\rm sp}=5$ ensemble shown in
    Figure~\ref{fig:epsilon5_lines}. {\sl Left}: Including the \bicep\
  data and {\sl Right}: excluding the \bicep\ data. In both cases the
  bottom plot shows the potentials re-centred around the value of
  $\phi$ at which the pivot scale $k=0.05$ Mpc$^{-1}$ is exiting the
  horizon. The origin of the suppression in scalar power is the mild, transient
  steepening of the potential around $(\phi-\phi_0)\sim 0.4$.}
  \label{fig:pot5}%
\end{figure*}

\subsection{Inflaton potential}

Each of the splined $\epsilon$ trajectories can be related to a scalar
potential $V(\phi)$ using~(\ref{potential}) and~(\ref{eq:phi}). This
assumes that inflation is driven by a single scalar with $\phi$
evolving monotonically in time. The relation does not depend on any
slow-roll assumption explicitly but there is an implicit dependence on
the first order assumption made in relating $\epsilon(N)$ to the
primordial power spectra in order to compare with the data.

Figure~\ref{fig:pot5}  shows the ensemble of potentials for the
1-$\sigma$ range of $\epsilon$ trajectories in the $N_{\rm sp}=5$
run for the case with and without \bicep\ respectively. The potentials are plotted against the value of $\phi$ centred on
$\phi_0$ - the value of $\phi$ when the largest mode is exiting the
horizon, and also centred on $\phi_\star$ - the value of $\phi$ when
the pivot scale is exiting the horizon. The latter shows how the
\bicep\ data leads to a tightly constrained normalisation {\sl and}
shape of the potential in the region around the pivot scale. The
former shows more clearly the shapes of each trajectory.

The origin of the suppression feature can be seen in the potential
shapes constrained by \bicep\ as a mild, transient steepening in most
of the best-fitting potentials in the initial stages as  the field
rolls down hill. In this transient region the field velocity increases
temporarily which results in an enhanced value of $\epsilon$ and
suppressed scalar power. The tensor spectrum is mostly unaffected as
it only depends on the integral of $\epsilon$. The total field
displacement is also tightly constrained when including \bicep\ since
the overall gradients of the curves are similar. This is in contrast
to the no \bicep\ case where a much greater range in gradients and
curvature of the potential is allowed resulting in very different
overall displacements in $\phi$ required to cover the observable
window.

\section{Discussion}\label{sec:disc}

We have introduced a method for obtaining posterior distributions in
parameters describing the acceleration of the background during
inflation. The posteriors are obtained by MCMC exploration of a data
likelihood. The method successfully fits the data and the
parametrisation is favoured by the \planck+\wwp+\bicep\ combination of
CMB data over traditional power law parametrisation of the primordial
spectra.

The method can also be considered as a procedure for the
reconstruction of the primordial spectra as we have shown by obtaining
derived distributions in $P_s(k)$. The tensor mode spectrum can also
be obtained but we have omitted it here as it does not show any
surprising feature. In the scalar case we have shown that the well
known suppression feature driven by the \bicep\ {\sl vs} \planck\ tension
is also present in this reconstruction and appears to be
significant. It is difficult to quantify the significance precisely
without comparing to a full distribution of allowed power law models
but, from our analysis, it is clear that the power law model is
incompatible at least at the 95\% confidence level - if not higher.

As a reconstruction method, our procedure can be compared with other
non-parametric methods such as those used in Section~7 of
\cite{planck_inf}. One advantage of the parametrised acceleration
method used here is that it imposes a physically justified smoothness
constraint on the spectrum. This is due to the monotonicity of $H$ and the
requirement that $\epsilon<1$ and avoids the problem of over-fitting
of the noise in the data which is evident in the structure obtained in
direct reconstruction methods.

The method can also be used to reconstruct the form of the inflaton
potential $V(\phi)$ under the specific assumptions that there is a single
inflaton driving the acceleration. In this case too, we have shown how to
obtain a posterior distribution in the possible shape of the potential
under these assumptions. The feature seen in the scalar power spectrum
is also present in the form of a subtle transient in the gradient of the
best-fitting potentials in the relevant range in $\phi$. This method
can also be compared to reconstruction methods that either parametrise
the potential as a Taylor series around a pivot point (see for example
Section~6 of \cite{planck_inf}) or Hamilton-Jacobi methods (see for
example \cite{contaldi_planck}).

If the \bicep\ result is confirmed by further observations it will
motivate the careful study of polarisation on large scales. This will
allow us to obtain as much information as we can from the tensor modes
and methods such as the one used here will be important tools in the
quest to understand what the data says about the theory of
inflation. Further improvements in the method itself are also
possible. A more sophisticated parametrisation of the acceleration
that results in less correlated variables should be developed. The
function $\epsilon(N)$ could also be expanded on a basis that has
better analytic properties than a cubic spline. This would be crucial
in order to extend the method to higher orders in slow-roll or to
employ exact solutions for the perturbations. This work is left for
future studies.

\begin{acknowledgements}
  We acknowledge useful discussions with J. Richard Bond, Jonathan
  Horner, and Marco Peloso. We thank the hospitality of the
  Perimeter Institute where some of this work was carried out.
\end{acknowledgements}

\end{document}